\documentclass[a4paper]{article}
\usepackage{amsmath,amssymb}
\usepackage[dvipdfmx]{hyperref}
\usepackage[dvipdfmx]{graphicx}
\usepackage{enumerate}

\title{Extension of the constant exchange probability method
to multi-dimensional replica exchange Monte Carlo
applied to the tri-critical spin-1 Blume-Capel model}
\author{Kenji Kimura and Saburo Higuchi \\
  Department of Applied Mathematics and Informatics,\\
  Ryukoku University, Otsu, Shiga 520-2194, Japan
}

\begin{document}
\maketitle

\begin{abstract}
In replica exchange Monte Carlo (REM), 
tuning of the temperature set and the exchange scheduling are crucial in improving the accuracy and reducing calculation time.
In multi-dimensional simulated tempering, the first order phase transition is accessible. Therefore it is important to 
study the tuning of parameter set and the scheduling of exchanges
in the parallel counterpart, the multi-dimensional REM.
We extend Hukushima's constant exchange probability method
to multi-dimensional REM for the parameter set.
We further propose a combined method to use this set and the Bittner-Nu{\ss}baumer-Janke's $\text{PT}_{\tau}$ algorithm for scheduling.
We test the proposed method in two-dimensional spin-$1$ Blume-Capel model and find that it works efficiently, including the vicinity of the first order phase transition.
\end{abstract}

\section{Introduction}
\label{sec:introduction}
Replica exchange Monte Carlo (REM) or parallel tempering is a well used method for Monte Carlo (MC) simulation \cite{hukushima1996exchange, marinari1997numerical}.
To enhance the efficiency of sampling of MC simulations with the method, it is necessary to tune the temperature set and the schedule of replica exchanges. 
For the former, Hukushima's constant exchange probability method \cite{hukushima1999prob} is one of the oldest and the most transparent one. Many authors consider other methods from various viewpoints\cite{katzgraber2006feedback, araki2013adaptive, vogel2015towards, bittner2008make,ballard2009replica}.

Multi-dimensional replica exchange Monte Carlo \cite{sugita2000multidimensional, fukunishi2002hamiltonian} with multiple coupling constants is a straightforward extension of REM to explore larger phase spaces.
Multi-dimensional simulated tempering \cite{mitsutake2009from} and simulated tempering and magnetizing \cite{nagai2012simulatedtempering, nagai2013simulated} are closely related methods. 

In systems having the first order phase transition, it has been known that replicas above and below the critical temperature hardly mix in uni-dimensional REM.
In ref.\cite{nagai2012simulatedtempering}, however, it is reported that one can study the vicinity of the first order phase transition by connecting the two sides of the transition line through the high temperature region.  It is naturally expected that the same mechanism works in the multi-dimensional REM. Therefore it is important to have an efficient implementation of it. Namely, we aim to study the efficient parameter set assignment to replicas and the schedule of replica exchange.

In this article, we consider multi-dimensional REM and extend Hukushima's constant exchange probability method. We propose to use it in combination with the Bittner-Nu{\ss}baumer-Janke's $\text{PT}_{\tau}$ algorithm for the scheduling. We test it in the spin-1 Blume-Capel model\cite{blume1966theory, capel1966on} and show its efficiency in the presence of the first order phase transition line.

\section{Multi-dimensional replica exchange Monte Carlo}
\label{sec:mdrem}
Let us consider a system with the following Hamiltonian
\begin{equation}
\mathcal{H}(x,J_1,\ldots,J_K) = \sum_{k=1}^K J_k E_k(x).
\label{hami_m1}
\end{equation}
where $x$ is a configuration, $J_k$ are coupling constants and $E_k$ are energy operators. 

To investigate the phase space $\{(J_1,\ldots,J_K)\}$ with uni-dimensional REM, we choose a finite set of $\vec{J}=(J_1,\dots,J_K)$ and for each of them we run a REM simulation with a set of replicas having temperatures $\{\beta_m\}$.
In contrast, in multi-dimensional REM, we set the temperature to unity and consider an extended system consisting of non-interactive $M_1 \times M_2 \times \cdots \times M_K$ replicas of the original system \eqref{hami_m1} with a set
$\vec{J}_{m_1,\cdots ,m_K} = ({J_1}_{m_1,\cdots ,m_K}, {J_2}_{m_1,\cdots ,m_K},\cdots, {J_K}_{m_1,\cdots ,m_K})$\, $(m_k=1,\ldots,M_k)$.

Hereafter, we restrict ourselves to the case $K=2$ for brevity of presentation and write $m=m_1, n=m_2$.
Generalization to $K>2$ case is straightforward.

A state of this extended system is specified by $M\times N$ configurations $\{ x\} = \{ x_{1,1}, x_{1,2}, \dots, x_{1,M}, x_{2,1}, \dots, x_{M,N}\}$.
We consider the probability distribution for the extended system given by
\begin{equation}
P(\{ x \} ; \beta ; \{ J_1 \} ; \{ J_2 \} ) = \prod_{m, n = 1}^{M,N} P(x_{m,n}, \beta ,{J_1}_{m,n}, {J_2}_{m,n}) = \prod_{m,n=1}^{M,N} \frac{\mathrm{e}^{- \beta \mathcal{H}(x_{m,n}, {J_1}_{m,n},{J_2}_{m,n})}}{Z(\beta, {J_1}_{m,n}, {J_2}_{m,n})},
\label{mrep_pd}
\end{equation}
where $\beta$ is inverse temperature, $\{ J_1\}$, $\{ J_2 \}$ are coupling constant sets and $Z$ is the partition function defined by
\begin{equation}
Z(\beta, {J_1}_{m,n}, {J_2}_{m,n}) = \sum_{x_{m,n}} \mathrm{e}^{-\beta  \mathcal{H}(x_{m,n}, {J_1}_{m,n}, {J_2}_{m,n})}.
\end{equation}

We exchange replicas with Metropolis algorithm.
Because $\beta$ always appears as a product $\beta J_k$ in \eqref{mrep_pd}, we fix $\beta$ among all replicas and exchange only $J_k$'s.
The transition probability for replica exchange process between the $(m, n)_\text{th}$ and the $(m^{\prime}, n^{\prime})_\text{th}$ replicas is given by
\begin{equation}
W(x_{m,n}, x_{m^\prime ,n^\prime}) = \min\left(1, \mathrm{e}^{-\Delta}\right),
\label{m_met}
\end{equation}
where we have introduced the cost function
\begin{equation}
  \begin{split}
\Delta = \beta \left\{\mathcal{H}(x_{m^\prime ,n^\prime}, {J_1}_{m,n}, {J_2}_{m,n}) + \mathcal{H}(x_{m,n}, {J_1}_{m^\prime ,n^\prime}, {J_2}_{m^\prime ,n^\prime})\right.\\ 
\left. - \mathcal{H}(x_{m,n}, {J_1}_{m,n}, {J_2}_{m,n}) - \mathcal{H}(x_{m^\prime ,n^\prime}, {J_1}_{m^\prime ,n^\prime}, {J_2}_{m^\prime,n^\prime}) \right\}.    
  \end{split}
\label{m_cost}
\end{equation}
There are many possible choices for implementation of the update.
Here, we adopt the following procedure:
\begin{enumerate}
\item for all $n$, attempt an exchange of each pair of replicas in the $m$ direction: $(m, n)_\text{th}$ and $(m+1, n)_\text{th}$ with odd $m$,
\item for all $n$, attempt an exchange of each pair of replicas in the $m$ direction:  $(m, n)_\text{th}$ and $(m+1, n)_\text{th}$ with even $m$, 
\item for all $m$, attempt an exchange of each pair of replicas in the $n$ direction:  $(m, n)_\text{th}$ and $(m, n+1)_\text{th}$ with odd $n$, 
\item for all $m$, attempt an exchange of each pair of replicas in the $n$ direction:  $(m, n)_\text{th}$ and $(m, n+1)_\text{th}$ with even $n$. 
\end{enumerate}
The four replica exchange steps above constitute one Monte Carlo step of multi-dimensional REM. 
One Monte Carlo step of local configuration update ($\text{MCS}_\text{local}$) is applied before each of the four steps.
Below, we count the number of replica exchange trials by the numbers of these steps performed.

\section{Method}
\label{sec:method}
We propose an iterative method to choose the constant set $\{\vec{J}\}$ to achieve the constant replica exchange probability in multi-dimensional REM. Further, we propose a combined method of the iteration and Bittner-Nu{\ss}baumer-Janke's $\text{PT}_{\tau}$ algorithm \cite{bittner2008make}.
Our iterative method is an extension of Hukushima's for REM \cite{hukushima1999prob} to the multi-dimensional REM.
The $\text{PT}_{\tau}$ algorithm has been proposed to maximize the number of round trips  at a fixed number of steps for a given parameter set in REM.

\subsection{Multi-dimensional constant exchange probability method}
We generalize Hukushima's constant exchange probability method\cite{hukushima1999prob} to multi-dimensional REM. We derive an iterative method from the cost function~\eqref{m_cost} for multi-dimensional replica system~\eqref{hami_m1}.

Because we work with $K=2$, Hamiltonian of the $(m,n)_\text{th}$ replica having $({J_1}_{m,n}, {J_2}_{m,n})$ is
\begin{equation}
 \mathcal{H}(x_{m,n}, {J_1}_{m,n}, {J_2}_{m,n}) = {J_1}_{m,n} E_1(x_{m,n}) + {J_2}_{m,n} E_2(x_{m,n}).
 \label{hami_m2}
\end{equation}
In this case, the cost function \eqref{m_cost} becomes
\begin{equation}
  \Delta=-\beta(\vec{J}_{m',n'}-\vec{J}_{m,n})\cdot(\vec{E}(x_{m',n'})-\vec{E}(x_{m,n})).
\end{equation}

The coupling constant set $\{\vec{J} \}= \{ \vec{J}_{1,1}, \vec{J}_{1,2}, \dots ,\vec{J}_{M,N}\}$ is placed on lattice points of a curved coordinate system in the coupling constant space. In the simplest case, we can think of constant spacing set
\begin{equation}
  \vec{J}_{m,n}=\vec{J}_0+ (m\delta_1   , n\delta_2),
\label{eq:const_spacing}
\end{equation}
where $\delta_1,\delta_2>0$ are spacings of a orthogonal lattice.

Our basis is that the cost function \eqref{m_cost} takes equal values for neighboring replica pairs. 
In the $m$ direction, the equality of $\Delta$ between pairs $((m\pm1,n),(m,n))$ implies
\begin{equation}
(\vec{J}_{m-1,n}-\vec{J}^\prime_{m,n})\cdot(\vec{E}_{m-1,n}-\vec{E}^\prime_{m,n})
=(\vec{J}_{m+1,n}-\vec{J}^\prime_{m,n})\cdot(\vec{E}_{m+1,n}-\vec{E}^\prime_{m,n}).
\label{mm_cost}
\end{equation}
The variable $\vec{E}_{m,n}$ above actually means 
the expectation values of internal energy $\langle \vec{E}(x_{m,n}) \rangle$.
Equivalently, eq.\eqref{mm_cost} follows by equating the average accept rate and making the approximation 
$\langle \mathrm{e}^{-\vec{J}\cdot\vec{E}(x)} \rangle = \mathrm{e}^{-\vec{J}\cdot\langle\vec{E}(x) \rangle}$.
In practice, $\vec{E}_{m,n}$ is evaluated by a preliminary short MC run and the reweighting.

Eq. \eqref{mm_cost} can be rewritten as
\begin{equation}
	( \vec{E}_{m+1,n} -  \vec{E}_{m-1, n} )\cdot \vec{J}^\prime_{m,n} = ( \vec{E}_{m+1,n} -  \vec{E}^\prime_{m,n}) \cdot \vec{J}_{m+1, n} -  ( \vec{E}_{m-1,n} -  \vec{E}^\prime_{m,n}) \cdot \vec{J}_{m-1, n}.
\label{mcprob1}
\end{equation}
Though we would like to solve \eqref{mcprob1} for unknowns $\vec{J}^\prime_{m,n} $, the number of conditions is insufficient to fix them.
Therefore, we impose an additional condition that the distance between $\vec{J}^\prime_{m,n}$ and the straight line connecting $\vec{J}_{m-1,n}$ and $\vec{J}_{m+1,n}$ is unchanged from the current $\vec{J}_{m,n}$:
\begin{equation}
  \vec{J}^\prime_{m,n}=\vec{J}_{m,n}+s\cdot (\vec{J}_{m+1,n}-\vec{J}_{m-1,n}),
 \label{mmm_cost}
\end{equation}
where $s$ is a real parameter. 
We can solve eqs.~\eqref{mcprob1},~\eqref{mmm_cost} to obtain
\begin{equation}
  s_{\mathrm{m}*}=\frac{ ( \vec{E}_{m+1,n} -  \vec{E}^\prime_{m,n})\cdot \vec{J}_{m+1, n} -  ( \vec{E}_{m-1,n} -  \vec{E}^\prime_{m,n}) \cdot\vec{J}_{m-1, n} -
 (\vec{E}_{m+1,n}-\vec{E}_{m-1,n})\cdot \vec{J}_{m,n}
}{
  (\vec{E}_{m+1,n}-\vec{E}_{m-1,n})\cdot(\vec{J}_{m+1,n}-\vec{J}_{m-1,n})
}.
\label{mcprob2}
\end{equation}
The denominator of the right hand side in eq.~\eqref{mcprob2} is non-zero generically. We have not met a situation where it vanishes when we apply this method. If it vanished, one could just skip the update of $\vec{J}_{m,n}$ and wait for $\vec{J}_{m\pm1,n}$ to be perturbed in the following steps.

By plugging \eqref{mcprob2} into \eqref{mmm_cost}, we obtain a formal solution to $\vec{J}^\prime_{m,n}$. It is no more than a formal solution because the expression for $s_{\mathrm{m}*}$ contains $\vec{E}^\prime_{m,n}$. Thus we iterate 
 \begin{equation}
\vec{J}^\prime_{m,n}=  \vec{F}(\vec{J}_{m,n})=\vec{J}_{m,n}+\left.s_{\mathrm{m}*}\right|_{\vec{E}^\prime=\vec{E}}\cdot (\vec{J}_{m+1,n}-\vec{J}_{m-1,n})
\label{gm}	
\end{equation}
and find the solution $\vec{J}^\prime_{m,n}$ as a fixed point of $\vec{J}^\prime_{m,n}=  \vec{F}(\vec{J}_{m,n})$. 

For the $n$ direction, 
the equality between the pairs $(m,n\pm1)$ leads to 
\begin{equation}
	( \vec{E}_{m,n+1} -  \vec{E}_{m, n-1} )\cdot \vec{J}'_{m,n} = ( \vec{E}_{m,n+1} -  \vec{E}^\prime_{m,n}) \cdot \vec{J}_{m, n+1} -  ( \vec{E}_{m,n-1} -  \vec{E}^\prime_{m,n}) \cdot \vec{J}_{m, n-1}
\label{mcprob1n}
\end{equation}
and then the recursion relation corresponding to eqs.\eqref{gm} and \eqref{mcprob2}.
\begin{gather}
\vec{J}^\prime_{m,n}=\vec{G}(\vec{J}_{m,n})=
\vec{J}_{m,n}+\left.s_{\mathrm{n}*}\right|_{\vec{E}^\prime=\vec{E}}\cdot (\vec{J}_{m,n+1}-\vec{J}_{m,n-1})
,\quad \\
  s_{\mathrm{n}*}=\frac{ ( \vec{E}_{m,n+1} -  \vec{E}^\prime_{m,n})\cdot \vec{J}_{m, n+1} -  ( \vec{E}_{m,n-1} -  \vec{E}^\prime_{m,n}) \cdot\vec{J}_{m, n-1} -
  (\vec{E}_{m,n+1}-\vec{E}_{m,n-1})\cdot \vec{J}_{m,n}
}{
  (\vec{E}_{m,n+1}-\vec{E}_{m,n-1})\cdot(\vec{J}_{m,n+1}-\vec{J}_{m,n-1})
}.
\label{gn}	
\end{gather}

In practice, we take the superposition of the two solutions. We add stabilization term to have the final form of the recursion relation 
\begin{equation}
	\vec{J}_{m,n} ^{t+1} = \frac{1}{2} (1-w)\times\frac12 \left(\vec{F}(\vec{J}_{m,n}^t) + \vec{G}(\vec{J}_{m,n}^t) \right)+ \frac12w \vec{C}^t + \frac{1}{2} \vec{J}_{m,n}^t,
\label{ggmn}
\end{equation}
where $t$ is the iteration step and $0\leq w<1$ is a parameter\footnote{One may think conditions~\eqref{mcprob1} and \eqref{mcprob1n} unambiguously fixes $\vec{J}'$. It turns out that, in practice,  this set of equations does not give rise to a recursion relation with a stable fixed point.}.
The first term in \eqref{ggmn} is  to enforce the replica pairs in the $m$ direction have equal exchange probability as well as the pairs in the $n$ direction. 
The second term
\begin{equation}
\vec{C}^t= \frac{1}{4}\left(\vec{J}_{m-1,n}+\vec{J}_{m+1,n}+\vec{J}_{m,n-1}+\vec{J}_{m,n+1}\right)
\label{gcorr}
\end{equation}
is to enforce that $\vec{J}^{t+1}_{m,n}$ stays within the quadrilateral formed by the four nearest neighbor replicas.
The lattice structure of replicas can collapse without this term. The third term is to enhance the convergence without changing the fixed point of the first and the second term.

If $\vec{J}_{m-1, n},\vec{J}_{m, n},\vec{J}_{m+1, n}$ are on straight line, there is a guarantee that eq.~\eqref{gm} has $\vec{J}_{m\pm1,n}$ as the periodic orbit of period $2$ and has a stable fixed point following Hukushima's argument\cite{hukushima1999prob}. If they are in a generic position, that argument does not apply.
This is the reason why we have to add stability terms in \eqref{ggmn} to enhance stability.

In our multi-dimensional constant exchange probability iterative method, the whole coupling constant set is divided into two checkerboard sublattices.
Using the iterative equation \eqref{ggmn}, one sublattice is updated while the other is kept fixed, and the process is repeated with the role of sublattices exchanged.
Our proposed method is to iterate these until all coupling constants converge. It produces a coupling constant set for which the replica exchange probability is approximately constant along each curves of replicas $m=$constant and $n=$ constant.

Note that we need special care for the boundary of the coupling constant lattice.
We adopt the following boundary condition.
Four corners of the rectangle shall be kept fixed.
On the boundary $m = 1$ and $M$, eq.\eqref{ggmn} shall be replaced with
\begin{equation}
\vec{J}_{m,n} ^{t+1} = \frac{1}{2} \left\{ \vec{J}_{m,n}^t + \vec{G}(\vec{J}_{m,n}^t) \right\},
\label{ggn}
\end{equation}
while for $n = 1$ and $N$, it shall be replaced with
\begin{equation}
\vec{J}_{m,n} ^{t+1} = \frac{1}{2} \left\{ \vec{J}_{m,n}^t + \vec{F}(\vec{J}_{m,n}^t) \right\},
\label{ggm}
\end{equation}
which is equivalent to Hukushima's method.

\subsection{Bittner-Nu{\ss}baumer-Janke's $\text{PT}_{\tau}$ algorithm}
A remarkable block structure that prevents replica exchanges near the second order phase transition has been found in the $t$-$\beta$ plot of replica trajectories by Bittner, Nu{\ss}baumer and Janke.
They have proposed a prescription for resolving this structure \cite{bittner2008make}.
It is to set the number of $\text{MCS}_\text{local}$  (denoted by $N_{\text{local}}$) between replica exchange attempts proportional to autocorrelation time $\tau(\beta)$ depending on the inverse temperature $\beta$. 
Though the computational time inevitably increases with the autocorrelation time,
this is by far more efficient than simply making $N_{\text{local}}$ uniformly large. For replicas with small $\tau(\beta)$, we can save computational time, especially in the parallel computational setting. We note that we can reduce the elapsed time by starting the calculation in descending order of $N_\text{local}$.
This $\text{PT}_\tau$ algorithm can be applied to multi-dimensional REM in a straightforward way.

\section{Application to spin-1 Blume-Capel model}
\label{sec:application}
We apply the proposed method to the spin-1 Blume-Capel model\cite{blume1966theory, capel1966on}.
Then, we verify that our proposed iterative method realizes constant exchange probability and increases the mixing of replicas.

\subsection{The model}
\begin{figure}[htb]
\begin{center}
\includegraphics[width=0.9\linewidth]{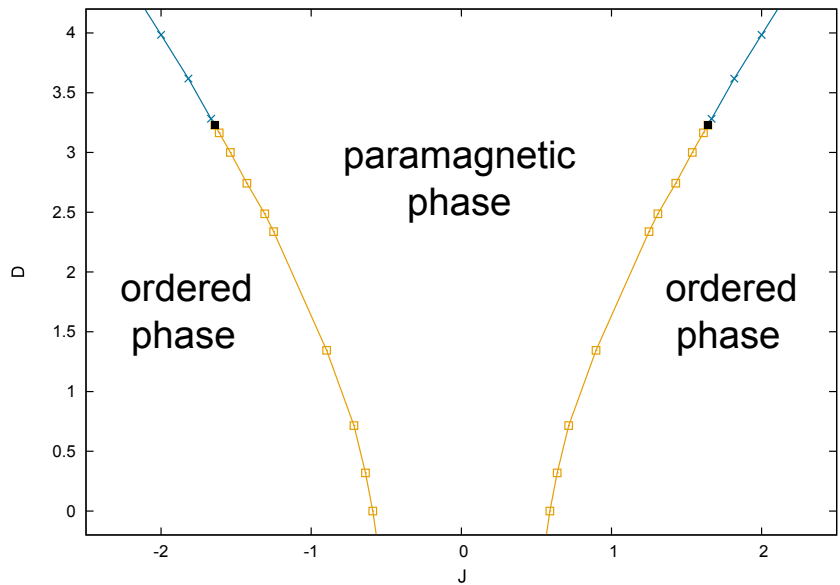}
\caption{Phase diagram of Blume-Capel model. Blue curves indicate the first order phase transition. Yellow curves indicate the second order phase transition. Filled black squares indicate tri-critical points.\cite{beale1986finite, xavier1998critical, silva2006wang,kwak2015first}}
\label{ph}
\end{center}
\end{figure}

Spin-1 Blume-Capel model is a generalization of the Ising model defined by the Hamiltonian
\begin{equation}
\mathcal{H}(\sigma,D,J) = -J\sum_{\langle ij\rangle} \sigma_i\sigma_j+ D \sum_i \sigma_i^2,
\label{bcmodel}
\end{equation}
where $J$ is a coupling constant, $D$ is the single-spin anisotropy parameter, and the spin variable $\sigma_i$ takes values $0, \pm 1$. The notation $\langle ij\rangle$ means all pairs of nearest-neighbor spins. We set $\beta=1$ without loss of generality below.
In two dimensions, this model has been studied well
and is known to have the first and the second order phase transition lines connected at the tri-critical point \cite{beale1986finite, xavier1998critical, silva2006wang,kwak2015first} (Fig. \ref{ph}). Therefore, it is suitable for testing our method.

\subsection{Proposed coupling constant set}
\label{sec:set}
For our proposed iterative method, this model falls into the $K=2$ case:
\begin{gather}
E_1 = \sum_{\langle ij\rangle} \sigma_i\sigma_j,\quad
E_2 =  \sum_i \sigma_i^2,\\
J_1 = J,\quad
J_2 = D.
\end{gather}
We set initial coupling constant set $\vec{J}$ on the sites of $M \times N$ rectangular lattice \eqref{eq:const_spacing}.
We apply our iterative method to the model \eqref{bcmodel} on the spatial $10\times10$ square lattice with the periodic boundary condition and obtain  a coupling constant set by the method\eqref{ggmn}. Then we perform multi-dimensional REM simulations with or without $\text{PT}_\tau$ algorithm and see whether the replica exchange is improved.

Our iterative method rearranges replicas in a given region in the coupling constant space with the boundary replicas constrained there. Therefore, the final constant exchange probability set could depend on the region considered. If the method is run on two overlapping regions, there is no guarantee that the set in the intersection agrees quantitatively or qualitatively. Moreover, it can depend on the initial lattice.

\begin{figure}[htb]
\begin{center}
\includegraphics[width=1\linewidth]{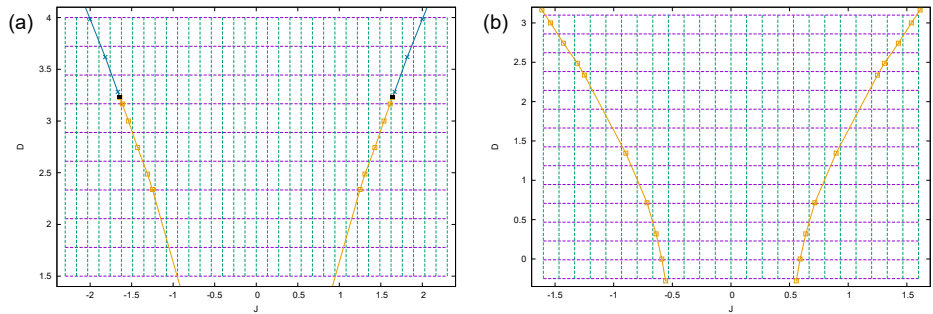}
\caption{Initial constant spacing set. Coupling constants of replicas are those at the intersections of vertical and horizontal lines. The replica exchange occurs between each connected replica pair. The yellow and the blue curves indicate the locations of the second and the first order phase transitions, respectively. The tri-critical point is indicated by filled black squares. (a) Region I. (b) Region II.}
\label{bc1_int}
\end{center}
\end{figure}

To inspect this situation closely, we test our method on the following two regions.
\begin{itemize}
\item Region I: $-2.3\leq J \leq +2.3, 1.5\leq D\leq 4$. Initial constant spacing lattice of $35 \times 10$ replicas (See Fig. \ref{bc1_int} (a))
\item Region II: $-1.6\leq J\leq +1.6, -0.25 \leq D \leq 3.1$. Initial constant spacing lattice of  $25 \times 15$ replicas (See Fig. \ref{bc1_int} (b))
\end{itemize}
Region I includes the first and the second order phase transition lines and the tri-critical point that connects the two. Region II includes only the second order phase transition line.
In both cases,  in order to respect the symmetry $J \leftrightarrow -J$, we make $J=0$ replicas stay on the $J=0$ line throughout the iteration by applying only eq.~\eqref{ggn} and not eq.~\eqref{ggm}.

\begin{figure}[h]
\begin{center}
\includegraphics[width=1\linewidth]{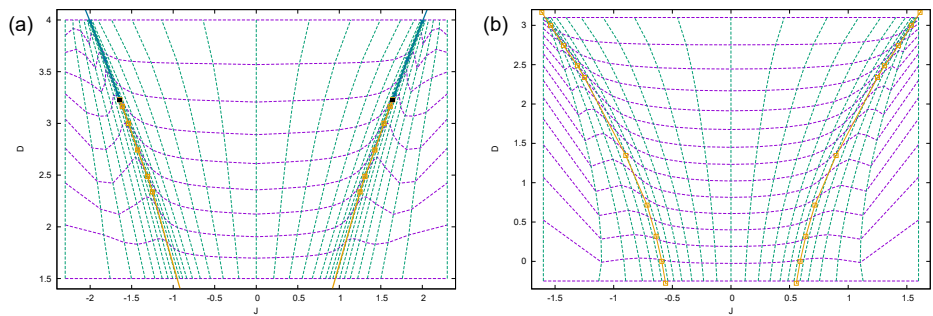}
\caption{Obtained constant exchange probability set. Replica exchange occurs on the dashing edges. The yellow and the blue curves indicate the locations of the second and the first order phase transition, respectively. The tri-critical point is indicated by filled black squares. (a) Region I. (b) Region II.}
\label{bc1_opt}
\end{center}
\end{figure}

For region I, we set $w = 0.1$ in \eqref{ggmn} because $w=0$ does not lead to convergence. For region II, we set $w=0.0$ and the lattice converges.
In both cases, our proposed iterative method makes 
the constant exchange probability set after around $2\times10^3$ steps.
As expected, the coupling constants concentrate near the transition line as Fig. \ref{bc1_opt}.

\begin{figure}[h]
\begin{minipage}{0.5\hsize}
\begin{center}
  \includegraphics[width=1\linewidth]{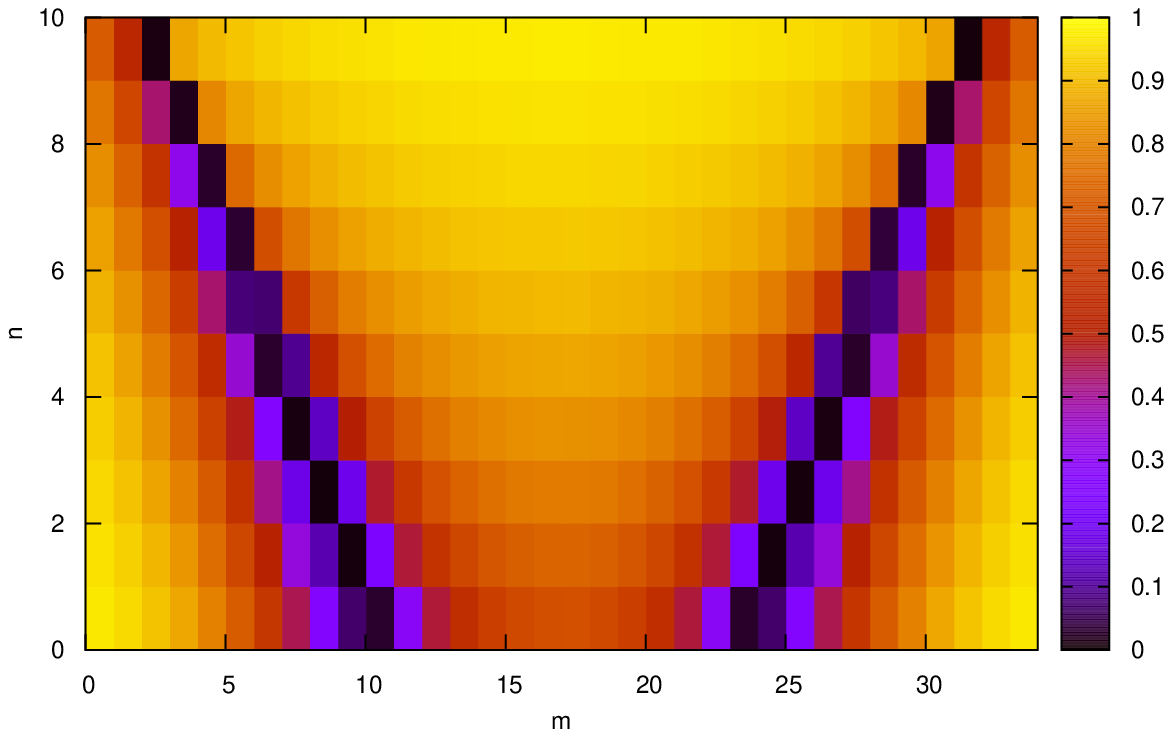}
\end{center}
\end{minipage}
\begin{minipage}{0.5\hsize}
\begin{center}
   \includegraphics[width=1\linewidth]{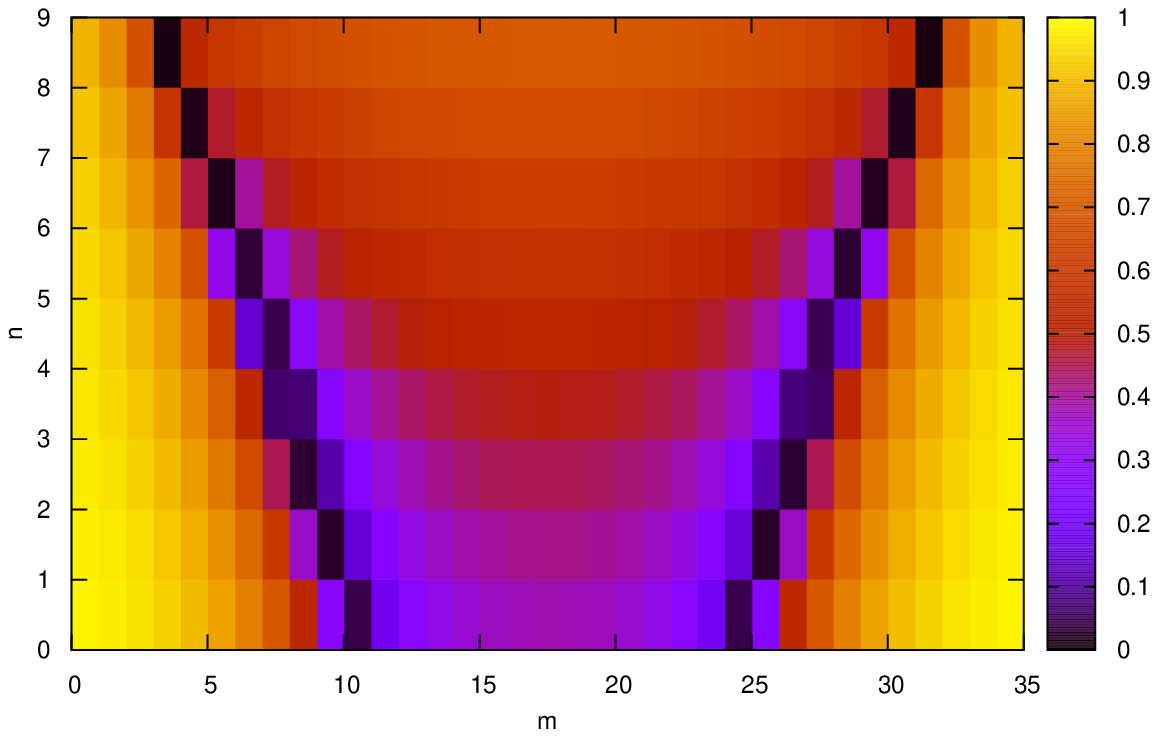}
\end{center}
\end{minipage}
\caption{Exchange probability of the constant spacing set in Region I  measured by $10^4$ exchange trials for each adjacent pair. (left) exchange probability of $m$ direction. (right) exchange probability of $n$ direction.}
\label{ex1_int}
\end{figure}

\begin{figure}[h]
\begin{minipage}{0.5\hsize}
\begin{center}
   \includegraphics[width=1\linewidth]{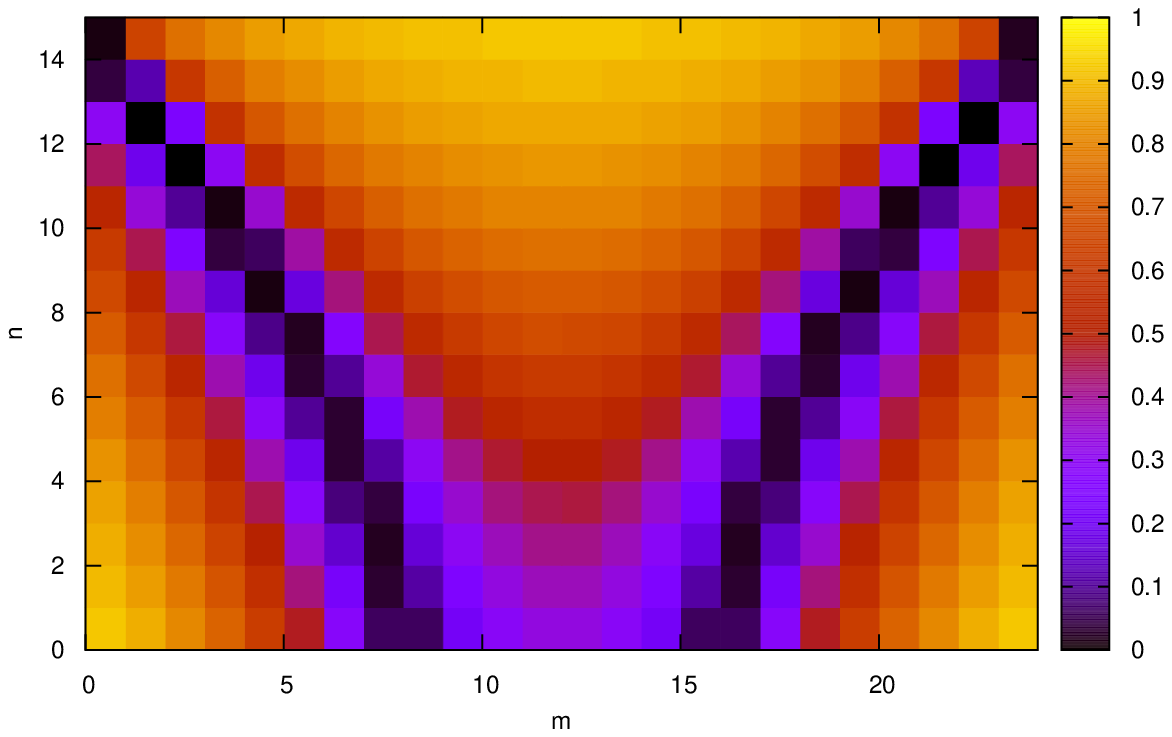}
\end{center}
\end{minipage}
\begin{minipage}{0.5\hsize}
\begin{center}
   \includegraphics[width=1\linewidth]{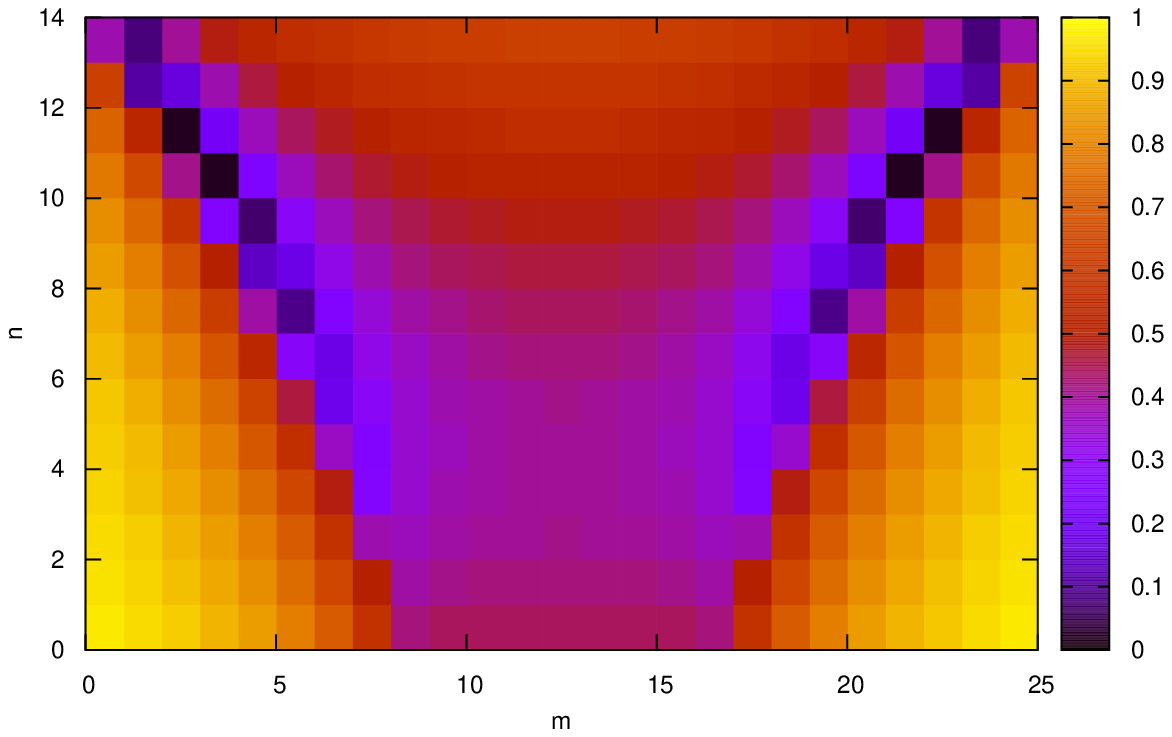}
\end{center}
\end{minipage}
\caption{Exchange probability of the constant spacing set in Region II. Same as Fig. \ref{ex1_int}.}
\label{ex2_int}
\end{figure}

\begin{figure}[h]
\begin{minipage}{0.5\hsize}
\begin{center}
   \includegraphics[width=1\linewidth]{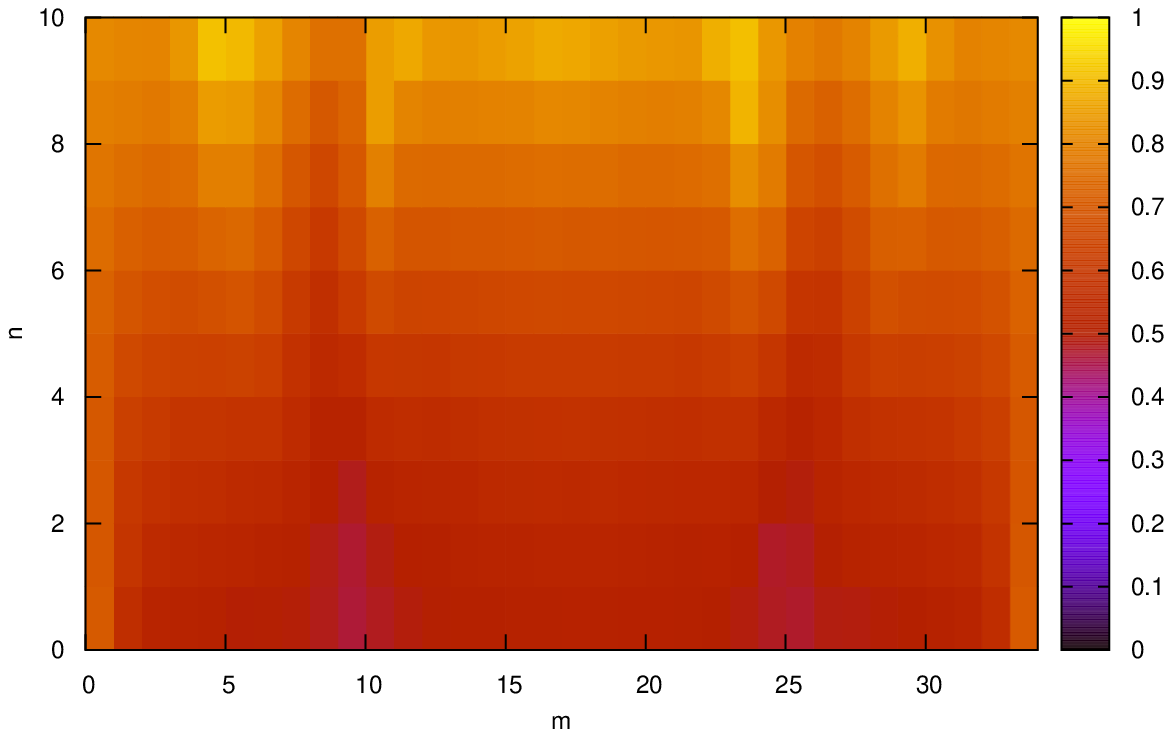}
\end{center}
\end{minipage}
\begin{minipage}{0.5\hsize}
\begin{center}
   \includegraphics[width=1\linewidth]{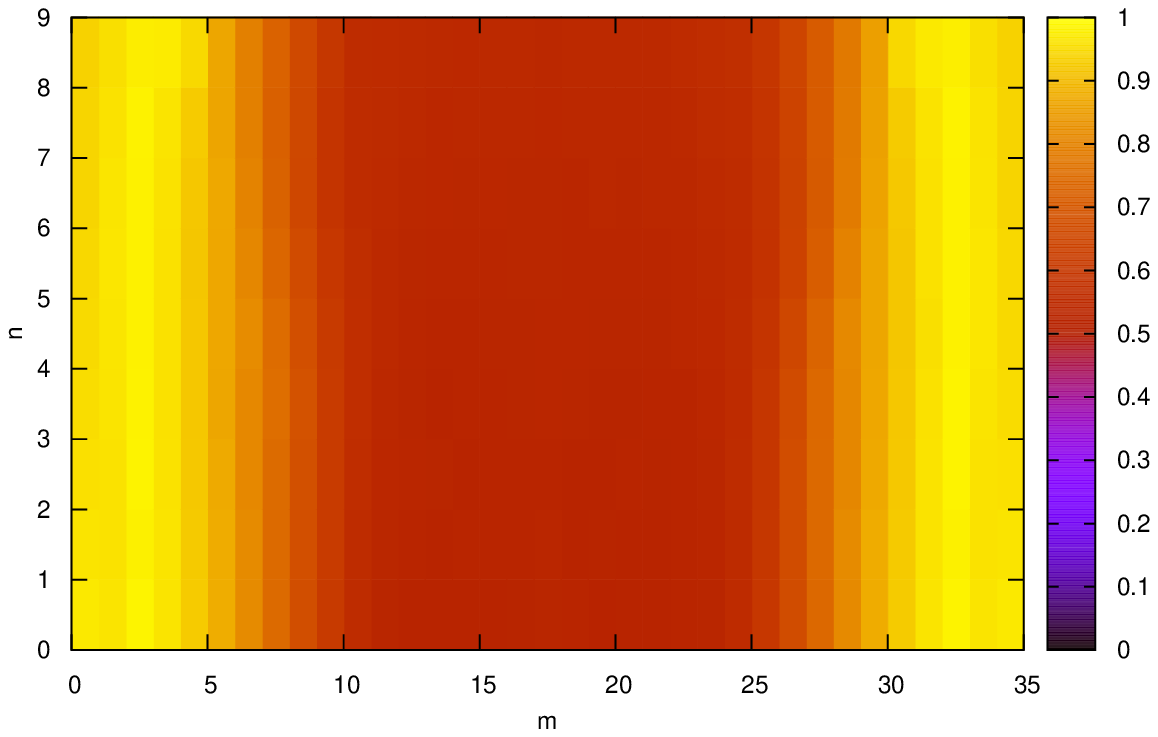}
\end{center}
\end{minipage}
\caption{Exchange probability of the constant exchange probability set in Region I  measured by $10^4$ exchange trials for each adjacent pair. (left) exchange probability of $m$ direction. (right) exchange probability of $n$ direction.}
\label{ex1_opt}
\end{figure}

\begin{figure}[htb]
\begin{minipage}{0.5\hsize}
\begin{center}
   \includegraphics[width=1\linewidth]{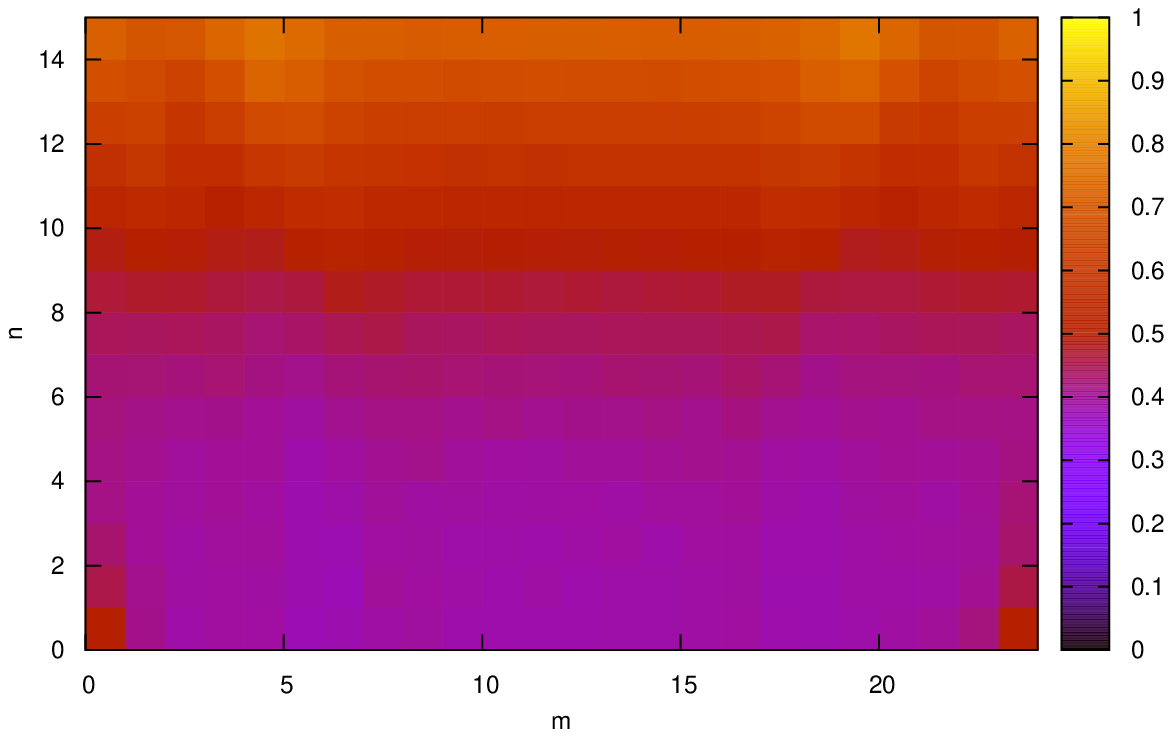}
\end{center}
\end{minipage}
\begin{minipage}{0.5\hsize}
\begin{center}
   \includegraphics[width=1\linewidth]{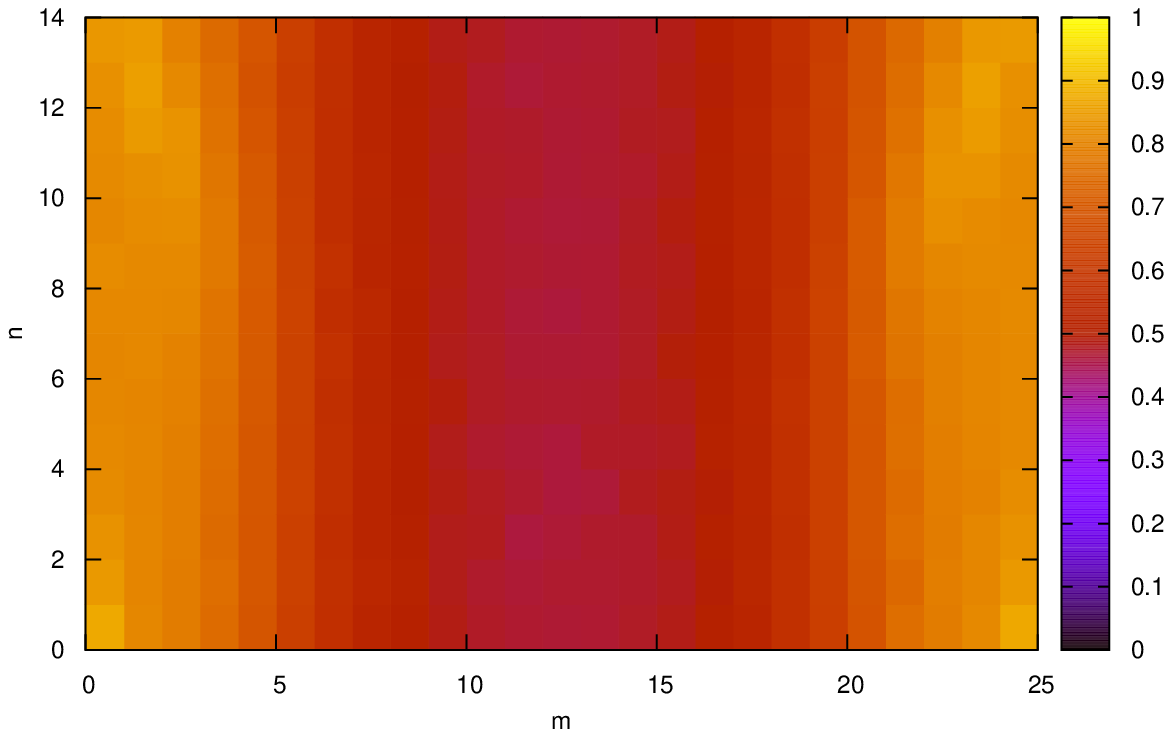}
\end{center}
\end{minipage}
\caption{Exchange probability of the constant exchange probability set in Region II. Same as Fig. \ref{ex1_opt}.}
\label{ex2_opt}
\end{figure}

Because of the existence of the transition line, for the constant spacing set, the exchange probability is extremely low there (Figs.~\ref{ex1_int} and \ref{ex2_int}). 
Due to the concentration of replicas and adaptive small spacings in the constant exchange probability set, the exchange probability between  $(m,n)_\text{th}$ and $(m+1,n)_\text{th}$ replicas becomes almost independent of $m$ for each $n$  (Figs.~\ref{ex1_opt} and \ref{ex2_opt}). The same holds for $n$ direction.

One notices that the exchange probability has weak dependence on $m$. It comes from the superposition of $m$ and $n$ directions in \eqref{ggmn} and the statistical error of the preliminary run to estimate the internal energy. In addition, for region I, there is an effect of $w>0$ term in \eqref{ggmn}.

We have tested with other initial lattices but qualitatively similar constant exchange probability set is obtained.

\subsection{The round trip time and replica trajectories}
We run multi-dimensional replica exchange Monte Carlo with the coupling constant set obtained in \S\ref{sec:set} and the $\text{PT}_\tau$ algorithm. 

We compare the round trip time for the constant spacing and constant exchange probability sets, with and without $\text{PT}_{\tau}$ algorithm.
When we employ $\text{PT}_{\tau}$ algorithm, we take  $N_{\text{local}}(J, D) \propto  \tau(J,D)$.
We use the exponential autocorrelation time $\tau_{\exp}(J,D)$ of the Hamiltonian operator. In the case without $\text{PT}_{\tau}$ algorithm. In those cases we perform $N_{\text{local}}(J, D) = 1$ local update after each of the four steps in the update procedure in \S\ref{sec:mdrem}. In both cases, the local configuration is updated with the local spin update Metropolis algorithm. 

It has been known that large replica exchange probability does not always mean that the all replica mix well. Moving in the coupling constant space is not a Markov process because each replica has its internal degrees of freedom as hidden variables. This situation arises as the notorious block structure in the $t$-$\beta$ plot.

In order to test the efficiency of replica exchange, one should take close look at the history of replica in the coupling constant space and see if a block structure arises or not\cite{bittner2008make}.  A quantitative measure of the mixing is the round trip time.
In our case, the round trip time ${\tau_\mathrm{rt}}_J$ is the time needed for a replica starting from $J=J_\text{min}$ to touch $J=J_\text{max}$ line and then come back to $J=J_\text{min}$.

The optimal round trip time is that for unbiased random walk of replicas on the coupling constant set without local configuration\cite{bittner2008make}. Because the hopping probability is anisotropic, we choose to compare with a random walk with prescribed hopping probability for constant exchange probability set. The net increase of ${\tau_\mathrm{rt}}_J$  compared to that for the random walk is explained by the effect of autocorrelation of local configurations.

\subsubsection{Region II}
We obtain the average round trip time ${\tau_{\text{rt}}}_J$ for several  methods as shown in Table \ref{rtt1}.
\begin{table}[h]
\begin{center}
\caption{Round trip time ${\tau_{\text{rt}}}_J$ in replica exchange steps defined in \S\ref{sec:mdrem}. These averages and errors are obtained from all replicas of 
$2\times10^5$ steps.}
\label{rtt1}
  \begin{tabular}{|c|c|}
\hline
   applied method ($N_{\text{local}}(J, D)$)& average ${\tau_{\text{rt}}}_J$ $\pm$ error\\
\hline
   constant spacing without $\text{PT}_{\tau}$& $19255.5	\pm 11695.3$\\
   constant spacing with $\text{PT}_{\tau}$ ($\tau_{\exp}$) &$7147.0\pm4259.4$\\
   constant exchange probability without $\text{PT}_{\tau}$& $6552.2\pm3902.1$\\
   constant exchange probability with $\text{PT}_{\tau}$ ($\tau_{\exp}$) &$2603.2\pm1508.3$\\
   constant exchange probability with $\text{PT}_{\tau}$ ($\frac{1}{2}\tau_{\exp}$) &$2951.5\pm1702.9$\\
   constant exchange probability with $\text{PT}_{\tau}$ ($2\tau_{\exp}$) &$2418.7\pm1406.8$\\
   constant exchange probability with $\text{PT}_{\tau}$ ($4\tau_{\exp}$) &$2359.1\pm1360.8$\\
   constant exchange probability, random walk& $2744.2\pm1641.1$\\
\hline
  \end{tabular}
\end{center}
\end{table}
The constant exchange probability set with $\text{PT}_{\tau}$ algorithm gives the least ${\tau_\mathrm{rt}}_J$, and it is almost as small as that in the random walk case.
This suggests that on our constant exchange probability set $\text{PT}_{\tau}$ algorithm works optimally.

To understand why the method works,
we examine the block structure \cite{bittner2008make} in $t$-$d$ plot.
We project 
replicas wandering in the $(J,D)$ space onto a single parameter $d$, the signed distance to the second order phase transition line.
We define $d$ to be positive for $(J,D)$ on the same side as $(J,D)=(0,0)$.

\begin{figure}[h]
\begin{minipage}{0.5\hsize}
\begin{center}
   \includegraphics[width=1\linewidth]{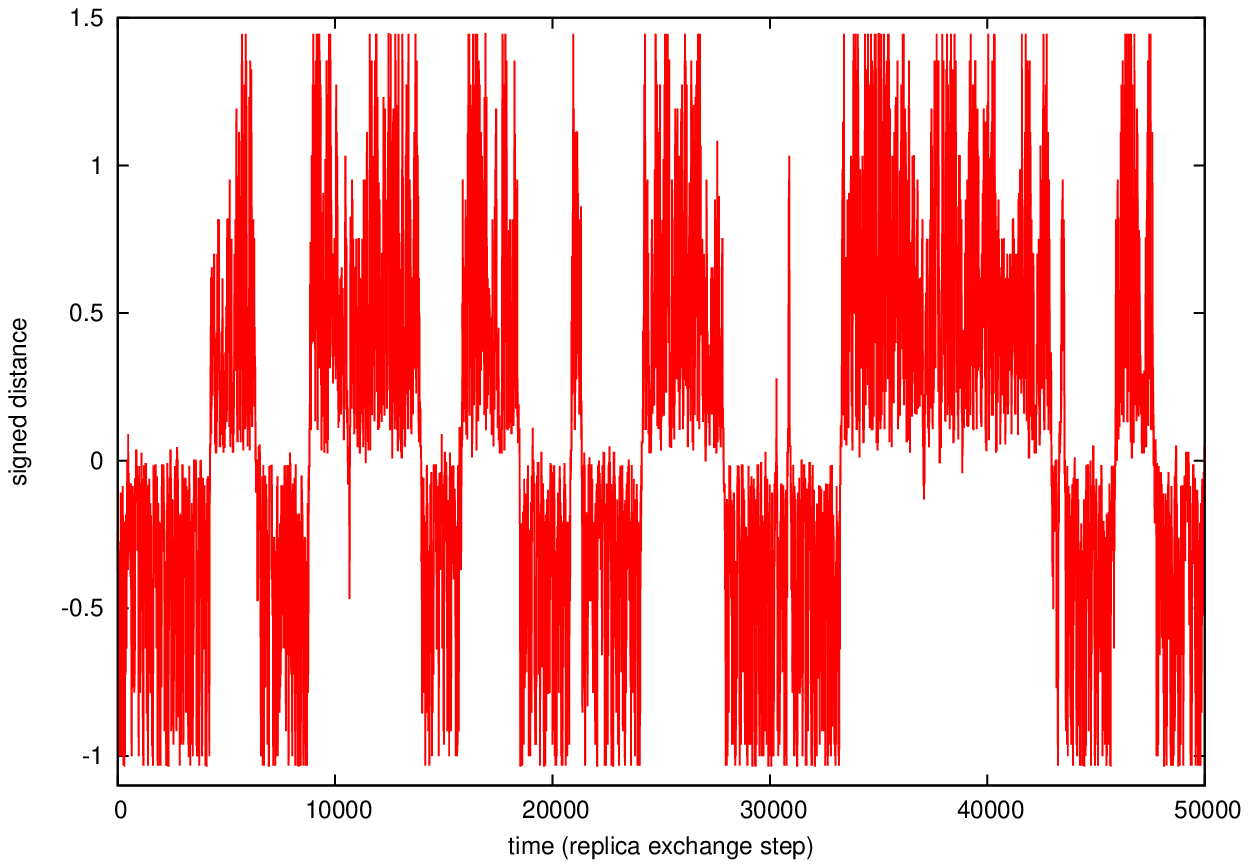}
\end{center}
\end{minipage}
\begin{minipage}{0.5\hsize}
\begin{center}
   \includegraphics[width=1\linewidth]{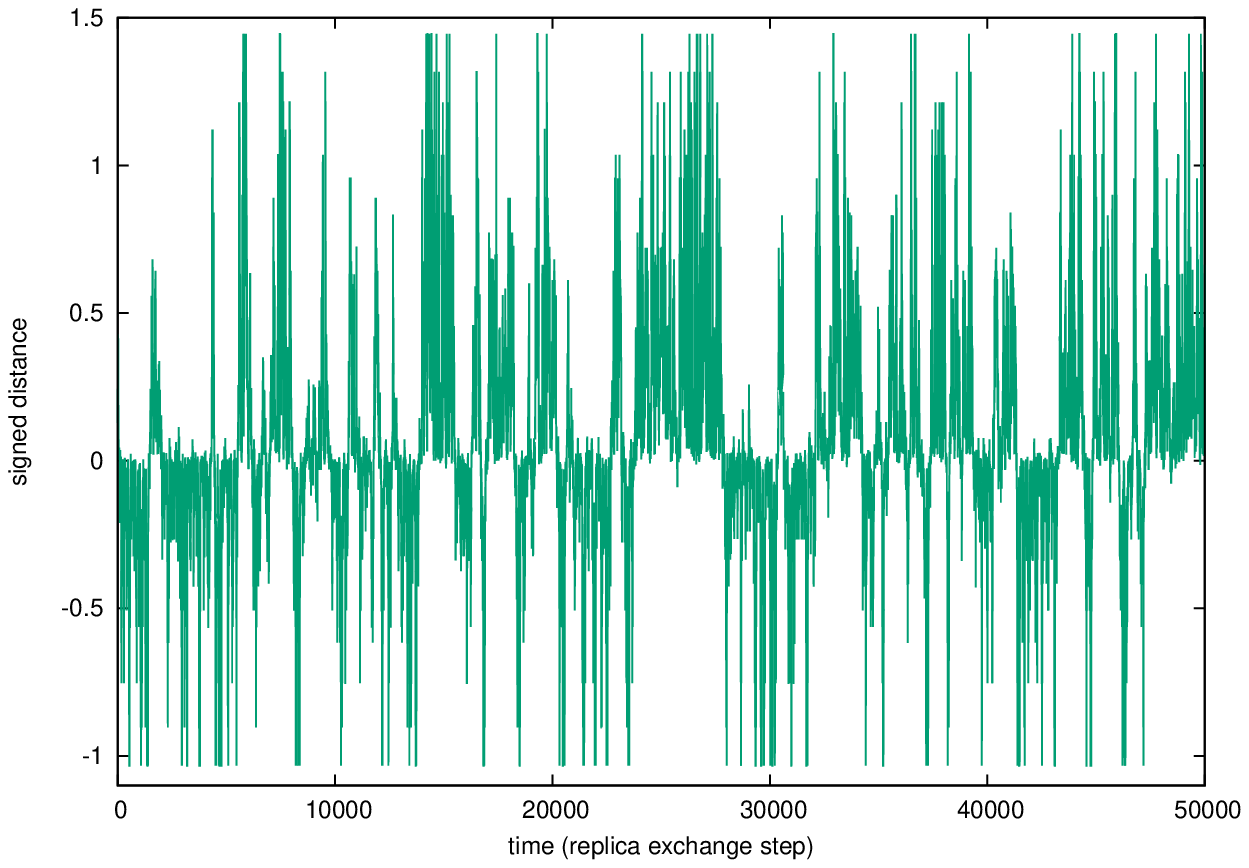}
\end{center}
\end{minipage}
\caption{History of a replica without $\text{PT}_{\tau}$ algorithm ($N_{\text{local}}(J, D)=1$). The horizontal axis is the number of replica exchange steps. The vertical axis represents the signed distance $d$. (left) constant spacing set. (right) constant exchange probability set. }
\label{dist_tran1}
\end{figure}

\begin{figure}[h]
\begin{minipage}{0.5\hsize}
\begin{center}
   \includegraphics[width=1\linewidth]{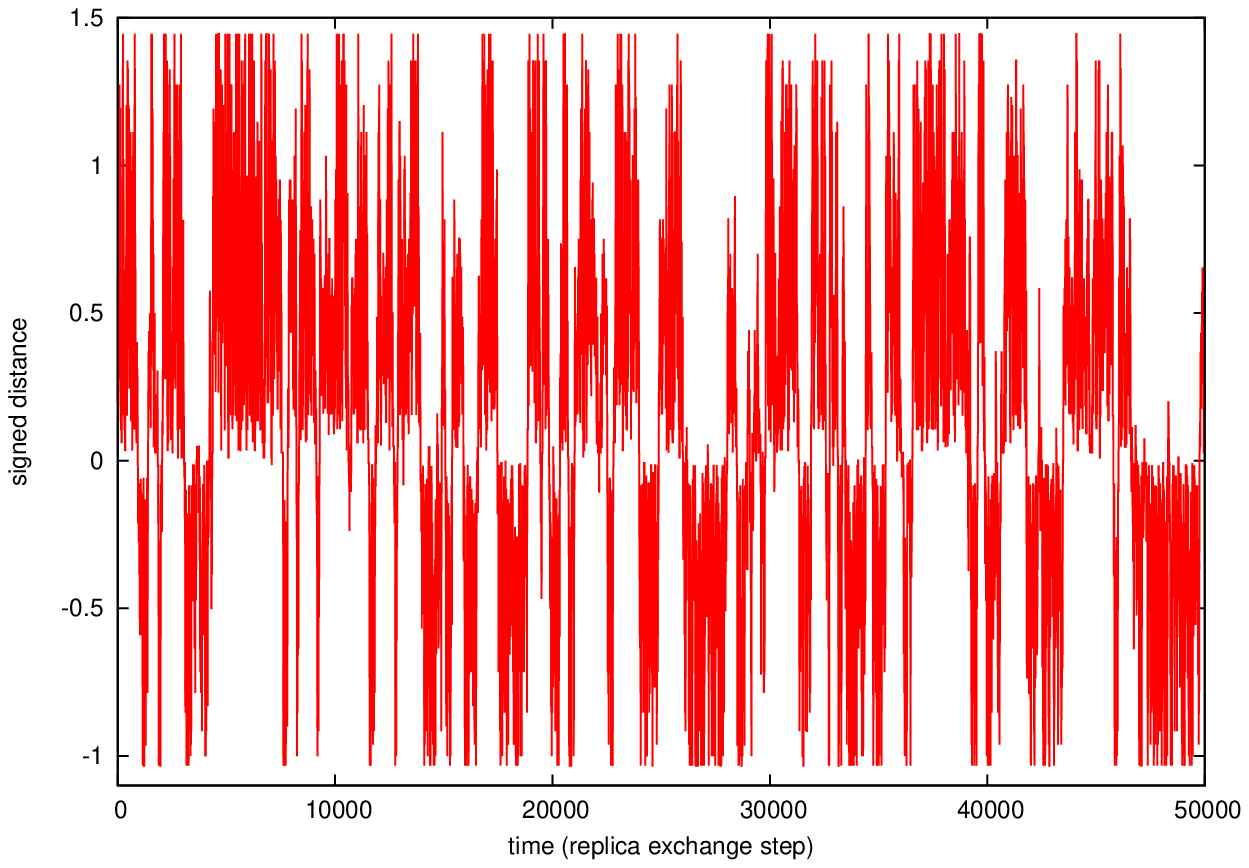}
\end{center}
\end{minipage}
\begin{minipage}{0.5\hsize}
\begin{center}
   \includegraphics[width=1\linewidth]{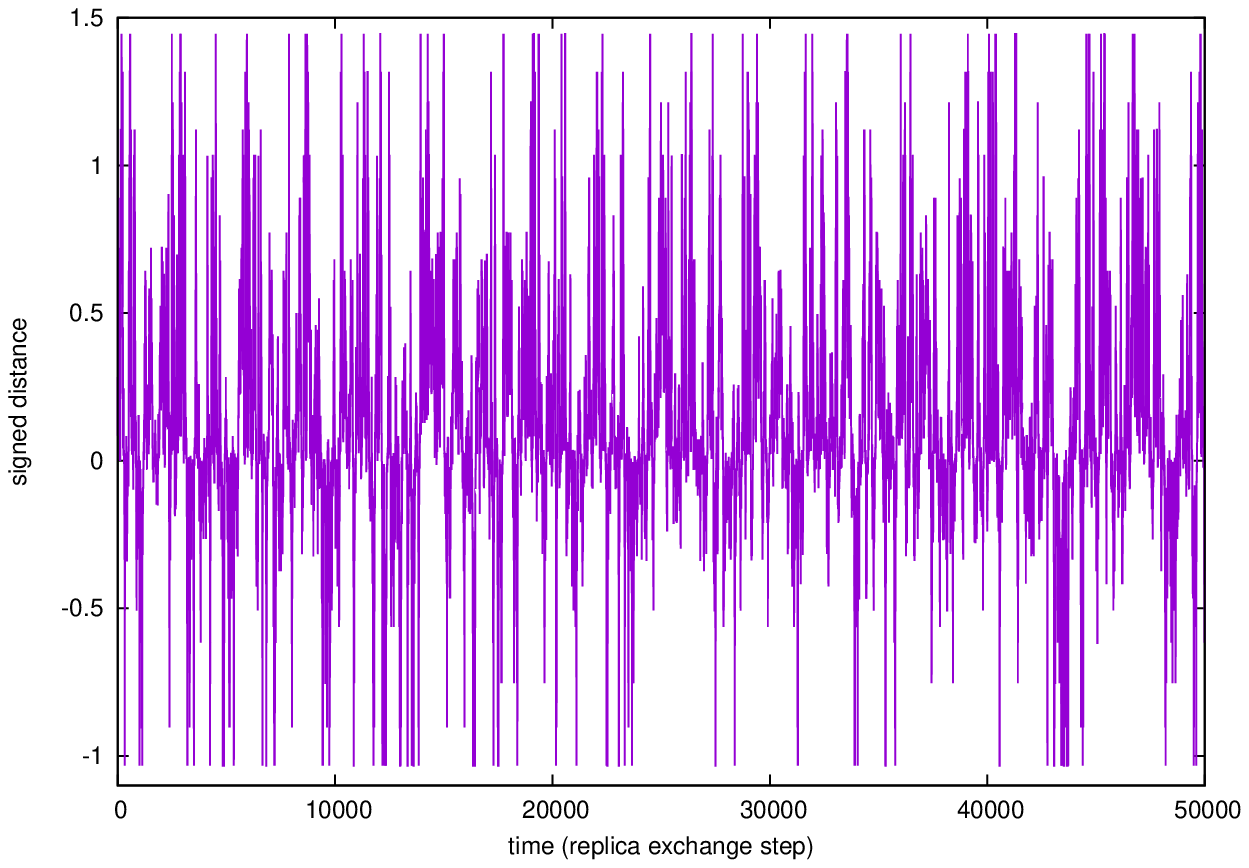}
\end{center}
\end{minipage}
\caption{History of a replica with $\text{PT}_{\tau}$ algorithm. Same as Fig. \ref{dist_tran1} for $N_{\text{local}}(J, D)=\tau_{\exp}$.}
\label{dist_tran2}
\end{figure}

In the case without $\text{PT}_{\tau}$ algorithm, transition of a replica makes the block structure, even if we use the constant exchange probability set as Fig. \ref{dist_tran1}. This means that, even if a replica happens to cross the line, there is large probability that its internal configuration does not change much and it jumps back to the original replica position at the next update.
Introduction of $\text{PT}_{\tau}$ algorithm resolves the block structure as seen in Fig. \ref{dist_tran2}.

\subsubsection{Region I}
Near the first order phase transition, the autocorrelation time is very large on a finite lattice.
We apply $\text{PT}_\tau$ algorithm with an upper limit $N_\mathrm{local}\lesssim \tau_{\exp}^{\mathrm{II}}$ instead of $N_\mathrm{local} > \tau_{\exp}^{\mathrm{II}} $ where $\tau_{\exp}^{\mathrm{II}}$ is that near the second order phase transition line.

We compare average round trip time ${\tau_{\text{rt}}}_J$ for several combinations of methods in Table \ref{rtt2}.
\begin{table}[h]
\begin{center}
\caption{Round trip time ${\tau_{\text{rt}}}_J$ in the replica exchange steps. These averages and errors are obtained from all replicas of $2\times10^5$ steps.}
\label{rtt2}
  \begin{tabular}{|c|c|}
\hline
   applied method (  $N_{\text{local}}(J, D)$ )& average ${\tau_{\text{rt}}}_J$ $\pm$ error\\
\hline
   constant spacing without $\text{PT}_{\tau}$& $65086.7\pm39897.1$\\
   constant spacing with $\text{PT}_{\tau}$ ($\tau_{\exp}$) &$12555.3\pm7767.2$\\
   constant exchange probability without $\text{PT}_{\tau}$& $15906.9	\pm9674.9$\\
   constant exchange probability with $\text{PT}_{\tau}$ ($\tau_{\exp}$) &$3198.1\pm1872.2$\\
   constant exchange probability, random walk & $2869.5\pm1741.9$\\
\hline
  \end{tabular}
\end{center}
\end{table}
The constant exchange probability set with $\text{PT}_{\tau}$ algorithm realizes the least ${\tau_\text{rt}}_J$. 
As expected, the result for the multi-dimensional REM is close to the random walk case in spite of the presence of the first order phase transition line.
The reason could be that the replica traverse the line through low $D$ region.

To examine this situation, we inspect a replica's trajectory in the two dimensional coupling constant space. To this end, we introduce the Euclidean angle $\theta$ formed between the tangent line of the first order phase transition line and the straight line connecting tri-critical point and the replica. In the region $J>0$, the angle $\theta$ shall be measured in the counter-clockwise direction around the tri-critical point.

\begin{figure}[h]
\begin{minipage}{0.5\hsize}
\begin{center}
	\includegraphics[width=1\linewidth]{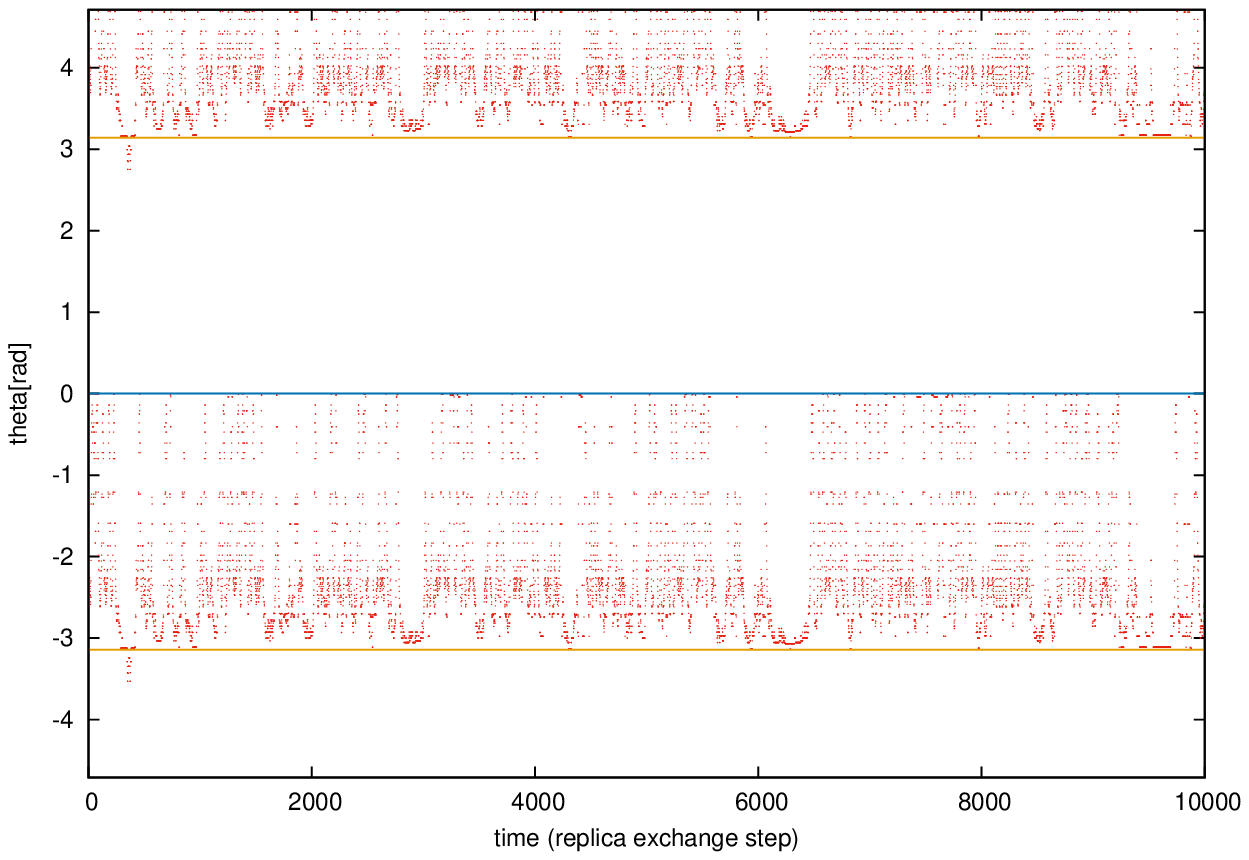}
\end{center}
\end{minipage}
\begin{minipage}{0.5\hsize}
\begin{center}
	\includegraphics[width=1\linewidth]{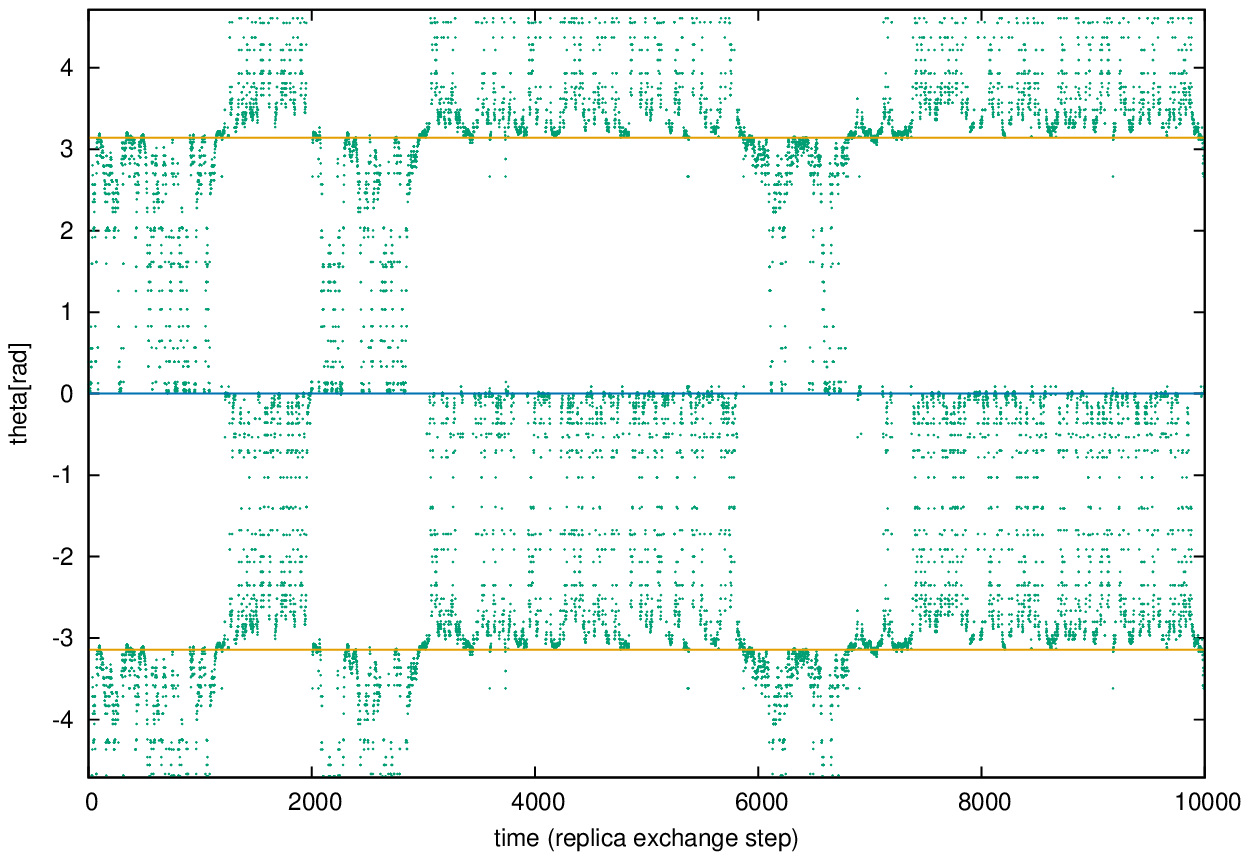}
\end{center}
\end{minipage}
\caption{History of a replica  without $\text{PT}_{\tau}$ algorithm ($N_{\text{local}}(J, D)=1$). The horizontal axis means time measured in the replica exchange steps. The vertical axis represents angle $\theta$. Blue horizontal line at $\theta = 0$ indicates the first order phase transition. Yellow horizontal lines at $\theta = \pm \pi$ indicate the second order phase transition. (left) constant spacing set. (right) constant exchange probability set.}
\label{angle1}
\end{figure}

\begin{figure}[h]
\begin{minipage}{0.5\hsize}
\begin{center}
	\includegraphics[width=1\linewidth]{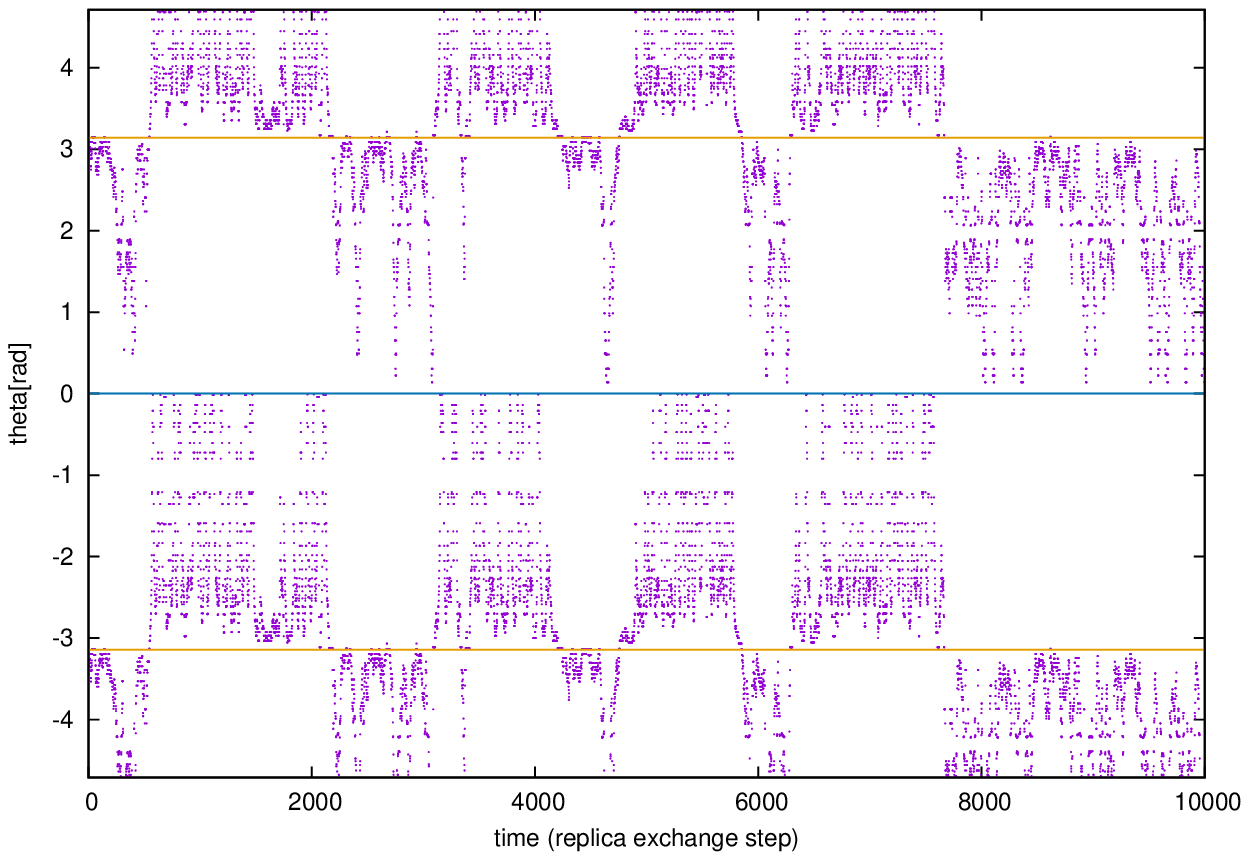}
\end{center}
\end{minipage}
\begin{minipage}{0.5\hsize}
\begin{center}
	\includegraphics[width=1\linewidth]{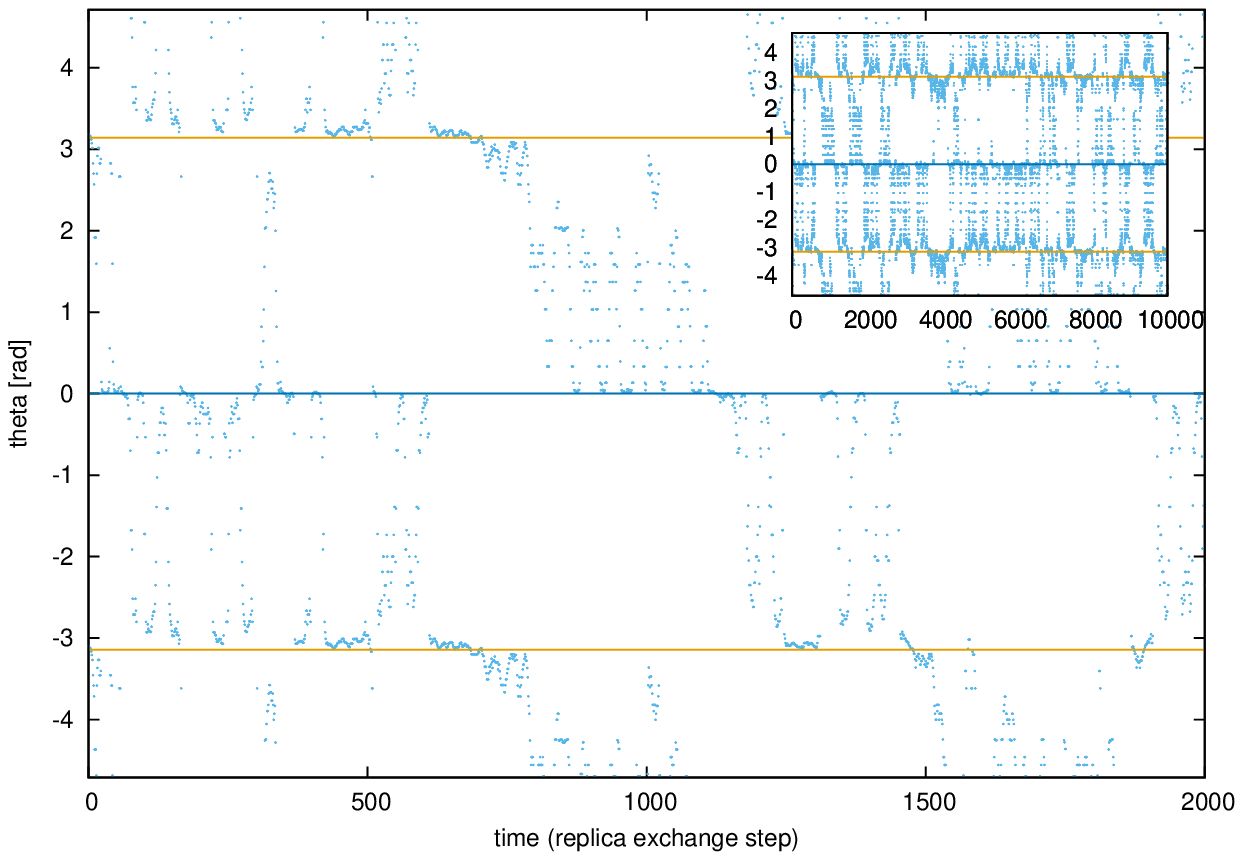}
\end{center}
\end{minipage}
\caption{History of a replica with $\text{PT}_{\tau}$ algorithm. Same as Fig. \ref{angle1} for $N_{\text{local}}(J, D)=\tau_{\exp}$. The right plot shows magnified view. (The inset shows the same time range as the left plot.) }
\label{angle2}
\end{figure}

We examine whether the block structure is observed in the parameter $\theta$.
In the case without $\text{PT}_{\tau}$ algorithm, trajectory of a replica in angle $\theta$ leads to the block  structure, even if we use constant exchange probability set as Fig. \ref{angle1}.
In the case with $\text{PT}_{\tau}$ algorithm, the block structure in the trajectory of replicas is resolved as shown in Fig. \ref{angle2}.

In the $t$-$\theta$ plots, 
$\theta=0\bmod 2\pi$ corresponds to the first order phase transition line,
while $\theta=\pi\bmod 2\pi$ corresponds to the second order one.
As shown in Fig. \ref{angle2}, many replicas failed to pass through the first order phase transition line but go around to cross the second order phase transition line in the low $D$ region.

\section{Discussion and Conclusions}
In this article, we have proposed the multi-dimensional constant exchange probability method and have proposed to use it in combination with $\text{PT}_{\tau}$ algorithm in multi-dimensional REM.
We have tested our method in spin-$1$ Blume-Capel model and have shown that this method improves the round trip time.
When we apply our combined method on a parameter region including the first and the second order phase transition, the replica exchange probability becomes almost constant for each direction. The round trip time is reduced because the replicas go around the tri-critical point and mix through the low $D$ region.

Not only REM but also Wang-Landau method\cite{wang2001efficient} and multicanonical algorithm \cite{berg1992multicanonical} can deal with such multi-dimensional coupling constant space \cite{zhou2006wang,valentim2015exploring,kwak2015first,mitsutake2009from}. Because each method has its advantage for measuring specific quantities of various models, it is desirable to compare the performance of our method with those of other methods in various situations.
In this work we have tested our method in a specific model at only small sizes. 
It is left for future work to test it for other models with $K>2$ coupling constants  and of larger spatial sizes.

It is known that one needs a set of replicas whose number is proportional to square root of the degrees of freedom to have large enough exchange probability\cite{fukunishi2002hamiltonian}. 
Though we have tested our method for a given numbers of replicas, the iterative method should work for arbitrarily large number of replicas in computational effort negligible in comparison to the main MC run. 
As for the main run, because we have shorter round trip time, 
we can expect that the small computational time would suffice to achieve fixed accuracy. It is also left for future work to determine how the total computational cost grows as the system becomes large.

\clearpage

\section*{Acknowledgments}
We are grateful to Shinji Iida and Junta Matsukidaira for discussions.

\bibliographystyle{unsrt}  
\bibliography{mc,bc}

\end{document}